\documentclass[sigconf]{acmart}
\usepackage{amsmath}
\usepackage{multirow}
\usepackage{subcaption}
\usepackage{booktabs}

\newcommand{\Real}{\mathcal{R}}

\AtBeginDocument{%
  \providecommand\BibTeX{{%
    \normalfont B\kern-0.5em{\scshape i\kern-0.25em b}\kern-0.8em\TeX}}}

\copyrightyear{2020}
\acmYear{2020}
\setcopyright{acmcopyright}

\acmConference[DLP-KDD 2020]{Proceedings of the 2nd International Workshop on Deep Learning Practice for High-Dimensional Sparse Data}{August 24, 2020}{San Diego, California, USA}
\acmBooktitle{Proceedings of the 2nd International Workshop on Deep Learning Practice for High-Dimensional Sparse Data (DLP '20), August 24, 2020, San Diego, California USA}
\acmPrice{15.00}
\acmDOI{10.1145/3306307.3328180}
\acmISBN{978-1-4503-6317-4/19/07}
\begin{document}

\title{FLEN: Leveraging Field for Scalable CTR Prediction}


\author{Wenqiang Chen}
\affiliation{%
  \institution{Data Intelligence, Meitu Inc}
  \city{Xiamen}
  \country{China}}
\email{wenqiang.cwq@gmail.com}

\author{Lizhang Zhan}
\affiliation{%
  \institution{Advertisement Recommendation Platform, Tencent Inc.}
  \city{Shenzhen}
  \country{China}}
\email{lizhangzhan@tencent.com}

\author{Yuanlong Ci}
\affiliation{%
  \institution{Data Intelligence, Meitu Inc}
  \city{Xiamen}
  \country{China}}
\email{cyl4@meitu.com}

\author{Minghua Yang}
\affiliation{%
  \institution{Data Intelligence, Meitu Inc}
  \city{Xiamen}
  \country{China}}
\email{jenniyang@meitu.com}

\author{Chen Lin}
\affiliation{%
  \institution{Xiamen University}
  \city{Xiamen}
  \country{China}}
\email{chenlin@xmu.edu.cn}
\authornote{Co-corresponding authors}

\author{Dugang Liu}
\affiliation{%
  \institution{Shenzhen University}
  \city{Shenzhen}
  \country{China}}
\email{dugang.ldg@gmail.com}
\authornotemark[1]
\renewcommand{\shortauthors}{Chen et al.}

\begin{abstract}
\textbf{C}lick-\textbf{T}hrough \textbf{R}ate (CTR) prediction systems are usually based on multi-field categorical features, i.e., every feature is categorical and belongs to one and only one field. Modeling feature conjunctions is crucial for CTR prediction accuracy. 
However, it usually requires a massive number of parameters to explicitly model all feature conjunctions, which is not scalable for real-world production systems. 

In this paper, we describe a novel \textbf{F}ield-\textbf{L}everaged \textbf{E}mbedding \textbf{N}etwork (\textbf{FLEN}) which has been deployed in the commercial recommender systems in Meitu and serves the main traffic. %a large-scale mobile photo-centric app.  
FLEN devises a field-wise bi-interaction pooling technique.
By suitably exploiting field information, the field-wise bi-interaction pooling layer captures both inter-field and intra-field feature conjunctions with a small number of model parameters and an acceptable time complexity for industrial applications.  
We show that some classic shallow CTR models can be regarded as special cases of this technique, i.e., MF, FM and FwFM.  
We identify a unique challenge in this technique, i.e., the FM module in our model may suffer from the coupled gradient issue, which will damage the performance of the model.
To solve this challenge, we develop \textbf{Dicefactor}: a novel dropout method to prevent independent latent features from co-adapting. %To our knowledge, this may be the first time field-aware feature conjunctions technique has been applied in real industrial systems.

Extensive experiments, including offline evaluations and online A/B testing on real production systems, demonstrate the effectiveness and efficiency of FLEN against the state-of-the-art models. In particular, compared to the previous version deployed on the system (i.e. NFM), FLEN has obtained $5.19\%$ improvement on CTR with $1/6$ of memory usage and computation time. 
\end{abstract}



\maketitle

\section{Introduction}
\textbf{C}lick-\textbf{T}hrough \textbf{R}ate (CTR) prediction is the task of predicting the probabilities of users clicking items or advertisements (ads). 
It is a critical problem in recommender systems and online advertising, which provide substantial revenue for Internet companies. 
As such, CTR prediction has attracted much attention from both academia and industry communities in the past few years~\citep{liu2017pbodl,chapelle2015simple,mcmahan2013ad,graepel2010web, zhou2018deep}.

The data in CTR prediction task is \textit{multi-field categorical data}, i.e., every feature is categorical and belongs to one and only one field. 
For example, feature ``gender=Female" belongs to field ``gender", feature ``age=24" belongs to field ``age" and feature ``item category=cosmetics" belongs to field ``item category". The value of  feature ``gender" is either ``male"  or``female.  Feature ``age" is discretized to several age groups: ``0-18", ``18-25", ``25-30", and so on. 
It is well regarded that, feature conjunctions are essential for accurate CTR prediction~\citep{cheng2016wide, covington2016deep, he2017neural, lian2018xdeepfm, guo2017deepfm}. 
An example of informative feature conjunctions is: age group ``18-25" combined with gender ``female" for item category ``cosmetics". It indicates that young girls are more likely to click on cosmetic products. 

Modeling sparse feature conjunctions has been continually improved and refined by a number of prior work. 
Most models follow the Factorization Machines (FM)~\citep{rendle2010factorization} and its inspired extensions because of its effectiveness and flexibility. 
For example, FFM~\citep{juan2017field} and FwFM~\citep{pan2018field,deng2020sparse} explicitly model field-aware feature interactions, NFFM~\citep{Yang2019Operation} combines FFM and MLP to capture operation-aware feature conjunctions with additional parameters, and FPENN~\citep{Liu2018Field} estimates the probability distribution of the field-aware embedding rather than using the single point estimation (the maximum a posteriori estimation).
Although these models have achieved promising results, \textit{the challenge in real-world online advertising or recommender systems is the strict latency limit at serving time and the scalability for high-dimensionality of features.} We need to predict hundreds of items for each user in less than 10 milliseconds.
The model complexity of FFM and FwFM is $O(N^2)$, where $N$ is the number of features. The drawback of directly applying FFM and FwFM in real-world applications is the dramatically increased use of computational resources, because any uniform increase in the number of features will cause a quadratic increase of computation.  
FFM is also restricted by space complexity, which further weakens its practicality. 
FFM-based deep models use additional parameters to capture non-linear high-order
feature conjunctions. However, increasing model complexity sometimes only marginally improve performance while leading to severe over-fitting problems~\citep{he2017neural}. In addition, these works usually consume huge memory, resulting in restricted scalability.

In this paper we describe a novel \textbf{F}ield-\textbf{L}everaged \textbf{E}mbedding \textbf{N}etwork (FLEN) which has been successfully deployed in the online recommender systems in Meitu, serving the main traffic. %is deployed in a large-scale mobile photo-centric app.  
FLEN devises a new operation in neural network modeling — Field-wise Bilinear Interaction (FwBI) pooling, to address the restriction of time and space complexity when applying field-aware feature conjunctions in real industrial system.
The field-wise bi-interaction pooling technique is based on the observation that features from various fields interact with each other differently.
We show that some classic shallow CTR models can be regarded as special cases of this technique, including FM~\citep{rendle2010factorization}, MF~\citep{Koren2009Matrix},  and FwFM~\citep{pan2018field}.
The combination of this technique and the traditional MLP layer constitutes our final model. The idea behind our model is to use multiple modules to extract feature interactions at different levels, and finally merge them to get a better representation of the interaction. 
The parameters in each part are only responsible for a certain level of feature interaction, which helps reduce the parameters of the model and speed up the training of the model.

As noted in previous work~\citep{qu2018product}, FM may cause the coupled gradient issue due to using the same latent vectors in different types of inter-field interactions, i.e. two supposedly independent features are updated in the same direction during the gradient update process.  
In FLEN, this issue is partly tackled by leveraging field information. 
We also propose a novel dropout method: Dicefactor to decouple independent features. 
Dicefactor randomly drops bi-linear paths (i.e. cross-feature edge in FM module of the field-wise bi-interaction pooling layer) to prevent a feature from adapting to other features. 
 
To demonstrate the effectiveness and efficiency of FLEN, we conduct extensive offline evaluations and online A/B testing. 
In offline evaluations, FLEN outperforms state-of-the-art methods on both a well-known benchmark and an industrial dataset consisting of historical click records collected in our system. 
Online A/B testing shows that FLEN enhances the CTR prediction accuracy (i.e. increases CTR by $5.19\%$) with a fraction of computation resources (i.e. $1/6$ memory usage and computation time), compared with the last version of our ranking system (i.e. NFM~\citep{he2017neural}). 

\section{Related work}
CTR prediction has been extensively studied in the literature, as online advertising systems have become the financial backbone of most Internet companies. 
Related literature can be roughly categorized into shallow and deep models. 

\subsection{Shallow Models}
Successful shallow models represent features as latent vectors. 
For example, matrix factorization (MF)~\citep{Koren2009Matrix} is successfully applied in recommender systems. It factorizes a rating matrix into a product of lower-dimensional sub-matrix, where each sub-matrix is the latent feature space for users and items. 
Factorization machines (FM)~\citep{rendle2010factorization} is a well-known model to learn feature interactions. In FM, the effect of feature conjunction is  explicitly modeled by inner product of two latent feature vectors (a.k.a. embedding vectors).
Many variants have been proposed based on FM. 
For example, \citet{rendle2011fast} proposed a context-aware CTR prediction method which factorized a three-way $<user, ad, context>$ tensor. \citet{oentaryo2014predicting} developed hierarchical importance-aware factorization machine to model dynamic impacts of ads. 

Field information has been acknowledged as crucial in CTR prediction. 
A number of recent work has exploited field information. 
 For example, Field-aware Factorization Machines (FFM)~\citep{juan2017field} represents a feature based on separate latent vectors, depending on the multiplying feature field. GBFM~\citep{cheng2014gradient} and AFM~\citep{xiao2017attentional} consider the importance of different field feature interactions. Field-weighted Factorization Machines (FwFM)~\citep{pan2018field} assigns interaction weights on each field pair. 
 
%FLEN is inspired by FwFM. However, the field-wise bi-interaction technique of FLEN generalizes and ensembles MF, FM and FwFM. 
The field-wise bi-interaction technique of FLEN is inspired by FwFM. It can be viewed as a special case of factorized FwFM in a computationally efficient manner. However, the field-wise bi-interaction technique of FLEN generalizes and ensembles MF, FM and FwFM.
Furthermore, shallow models are limited as they focus on modeling linear, low-order feature interactions. 
FLEN is capable of capturing not only low-order but also high-order, nonlinear interactions.

\subsection{Deep Models}
An increased interest in designing deep models for CTR prediction has emerged in recent years. 
The majority of them utilize feature bi-interactions. 
To name a few, NFM~\citep{he2017neural} generalizes FM by stacking neural network layers on top of a bi-interaction pooling layer. 
The architecture of DeepFM~\citep{guo2017deepfm} resembles with Wide\&Deep~\citep{cheng2016wide}, which also has a shared raw feature input to both its "wide" (i.e. for bi-interaction) and "deep" (i.e. for high-order interaction) components. 
DCN~\citep{wang2017deep} learns certain bounded-degree feature interactions. 
xDeepFM~\citep{lian2018xdeepfm}  improves over DeepFM and DCN by generating feature interactions in an explicit fashion and at the field-wise level.   
NFFM~\citep{Yang2019Operation} learns different feature representations for convolutional operations and product operations. 
However, the space complexity of NFFM and the time complexity of xDeepFM  restrict them from applying in industrial systems.
FGCNN~\citep{Liu2019Feature} leverages the strength of CNN to generate local patterns and recombines them to generate new features. Then deep classifier is built upon the augmented feature space.
PIN~\citep{qu2018product} generalizes the kernel product of feature bi-interactions in a net-in-net architecture.

The rest of literature learns the high-order feature interactions in an implicit way, e.g. PNN~\citep{qu2018product}, FNN~\citep{zhang2016deep}, DeepCrossing~\citep{shan2016deep}, and so on. Some tree-based methods~\citep{zhu2017deep,wang2018tem} combine the power of embedding-based models and tree-based models to boost explainability. One drawback of these approaches is having to break training procedure into multiple stages.
A recent work FPENN~\citep{Liu2018Field} also groups feature embedding vectors based on field information in deep neural network structure. 
It estimates the probability distribution of the field-aware embedding rather than using the single point estimation (the maximum a posteriori estimation) to prevent overfitting. 
However, as FPENN assigns several latent vectors to each field (i.e. one for a field which is not equivalent as the multiplying field), it requires much more model parameters than FLEN. 

\section{Model}\label{sec::model}
\begin{figure}
\centering\includegraphics[width=\columnwidth]{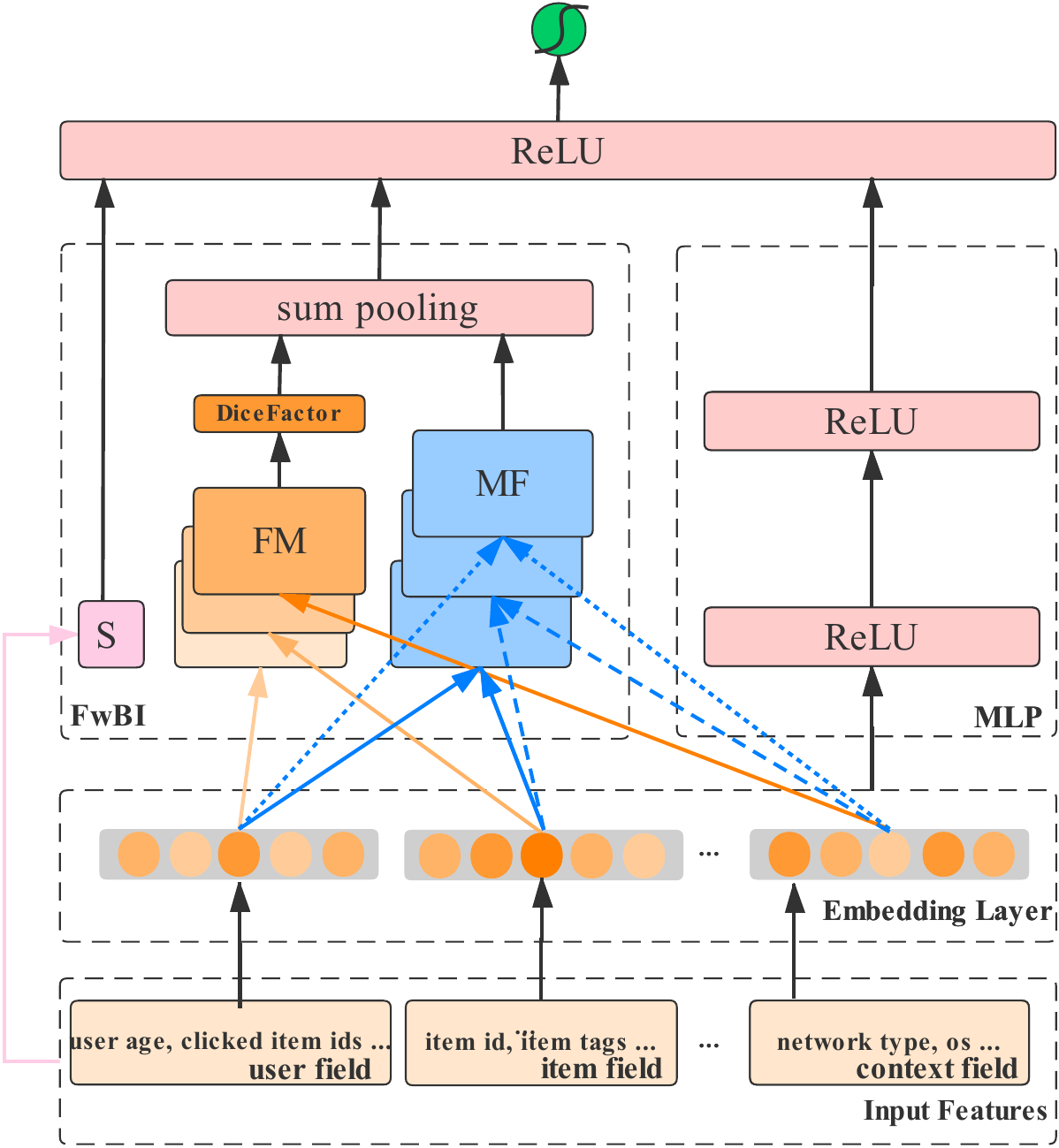}
\caption{Architecture overview of FLEN}
\label{fig:overview}
\end{figure}

An overview of the model architecture is illustrated in Figure~\ref{fig:overview}.
Let $\mathcal{X}=\{\mathbf{x}^1,\mathbf{x}^2,\cdots,\mathbf{x}^s,\cdots\}$ denote the set of samples, 
$\mathbf{x}^{s}=[\mathbf{x}_1^{s},\cdots, \mathbf{x}_N^{s}]$ denotes the $s$-th sample, and $\mathbf{x}_n^{s}\in\Real^{K_n}$ denotes the $n$-th categorical feature in the sample $\mathbf{x}^{s}$ (details described in Section~\ref{sec:feature}). 
Suppose FLEN takes $\mathbf{x}^{s}$ as input.
First, all features pass through an embedding layer, which outputs a concatenation of field-wise feature embedding vectors $\mathbf{e}^{s}=[\mathbf{e}_{1}^s,\cdots,\mathbf{e}_M^s]$, where $\mathbf{e}_m^s\in \Real^{K_e}$ denotes the $m$-th hierarchical fields (details described in Section~\ref{sec:embedding}).
Then the embedding vectors flow a Field-wise Bi-Interaction pooling layer (FwBI) and an MLP component. The field-wise bi-interaction pooling layer (Sec~\ref{sec:interaction}) consists of three submodules which capture all single and field-wise feature interactions (degree one or two), and outputs $\mathbf{h}_{FwBI}\in\Real^{K_{e}+1}$.
The MLP component captures non-linear, high-order feature interactions (details described in Section~\ref{sec:MLP}). The output of field-wise bi-interaction pooling layer and output of MLP component are concatenated to feed the last prediction layer (Section~\ref{sec:prediction}). 

We will show that previous CTR prediction models such as MF~\citep{Koren2009Matrix}, FM~\citep{rendle2010factorization} and FwFM~\citep{pan2018field} can be expressed and generalized under the proposed framework. 
Alternatively, the FwBI layer can be regarded as a combination of single feature, MF-based inter-field feature interactions and FM-based intra-field feature interactions. The above three parts together with the MLP layer that captures non-linear, high-order feature interactions form our model. The idea behind our model is to use multiple modules to extract feature interactions at different levels, and finally merge them to get a better representation of the interaction. The parameters in each part are only responsible for a certain level of feature interaction, which helps reduce the parameters of the model and speed up the training of the model. Furthermore,the parallel structures of FwBI allows the computational budget be distributed in a distributed environment.

Hereafter, unless stated otherwise, we use lower-case letters for indices, upper-case letters for universal constants, lower-case bold-face letters for vectors and upper-case bold-face letters for matrices, calligraphic letters for sets. We use square brackets to denote elements in a vector or a matrix, e.g. $x[j]$ denotes the $j-$th elements of $\mathbf{x}$. We will omit superscripts whenever no ambiguity results.
%lots need proof-reading

\subsection{Feature Representation}\label{sec:feature}

It is natural to represent categorical features as one-hot or multi-hot vectors. 
Suppose there are $N$ unique features and each feature $\mathbf{x}_n$ has $K_n$ unique values, we represent each instance as $\mathbf{x}=[\mathbf{x}_1,\cdots, \mathbf{x}_N]$, where  $\mathbf{x}_n\in\Real^{K_n}$, $\mathbf{x}_n[j]\in\{0,1\}, j=1,\cdots,K_n$. $\sum_{j=1}^{K_n}\mathbf{x}_n[j]=k$. Vector $\mathbf{x}_n$ with $k=1$ refers to one-hot encoding and $k>1$ refers to multi-hot encoding.
% A non-zero entry $x_n[j]=1$ indicates that the corresponding feature value in instance $x$ falls in the category $j$. 

Suppose there are $M$ fields, $F(n)$ denotes the field of feature $\mathbf{x}_n$, we organize the feature representations in a field-wise manner for complexity reduction. 
Specifically, $\mathbf{x}=concat(\mathbf{x}_1,\cdots,\mathbf{x}_M)$, where $\mathbf{x}_m$ is the concatenation of feature vectors in field $m$, i.e. $\mathbf{x}_m=concat(\mathbf{x}_n| F(n)=m)$. 
The organized instance is illustrated in the example below. Note that there is a hierarchical structure of fields. For example, ``item tags field" and ``item id field" belong to the more general ``item field". 
In practice, inspired by YouTube's work~\citep{covington2016deep}, we also classify features according to whether they describe properties of the item or properties of the user/context (namely user field, item field and context field as illustrated in Figure~\ref{fig:overview}) and achieve maximal performance enhancement and complexity reduction. A practically useful aspect of this hierarchical design is that it aligns with the intuition that the output of each hierarchical field is highly inter-correlated.
\begin{equation*}
    \underbrace{[\underbrace{0,1,0,...,0}_{\textit{age field}}]\ ... \ [\underbrace{1,0}_{\textit{gender field}}]}_{\textit{user field}}\ \underbrace{[\underbrace{0,1,0,...,0}_{\textit{item id field}}]\ [\underbrace{0,1,0,1,...,0}_{\textit{item tags field}}]}_{\textit{item field}}
\end{equation*} 

\subsection{Embedding Layer}\label{sec:embedding}
Since the feature representations of the categorical features are very sparse and high-dimensional, we employ an embedding procedure to transform them into low dimensional, dense real-value vectors. 
Firstly, we transform each feature $\mathbf{x}_n$ to $\mathbf{e}_n$.   
\begin{equation}
\mathbf{e}_n = \mathbf{V}_{n}\mathbf{x}_n, 
\end{equation}
where $\mathbf{V}_{n}\in\Real^{K_e \times K_n } $ is an embedding matrix for the corresponding feature that will be optimized together with other parameters in the network. Note that feature size can be various. 

Next we apply sum-pooling to $\mathbf{e}_n$ to obtain the field-wise embedding vectors. 
\begin{equation}
\mathbf{e}_m = \sum_{n|F(n)=m}\mathbf{e}_n
\end{equation}

\begin{figure}[htbp]
\centering\includegraphics[width=0.86\linewidth, scale=0.5]{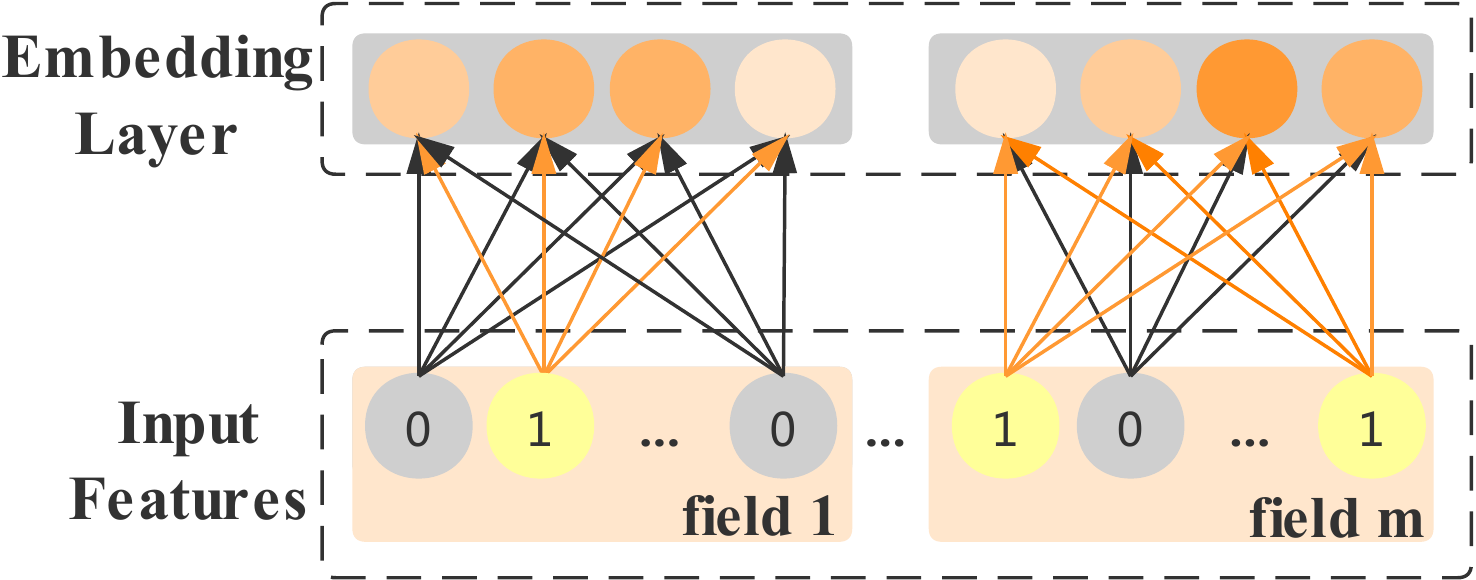}
\caption{Illustration of embedding layer with $K_e=4$.}
\label{fig:field_embedding}
\end{figure}

Finally, we concatenate all field-wise embedding vectors to build $\mathbf{e}$, as illustrated in Figure~\ref{fig:field_embedding}. 

\subsection{Field-wise Bi-Interaction Pooling Layer}\label{sec:interaction}

The field-wise bi-interaction pooling layer learns a mapping \\ $\Phi_{FwBI}(\mathbf{W}_{S},\mathbf{R},\mathbf{W}_{FwBI}): (\Real^{N},\Real^{MK_e})\rightarrow \Real^{K_{e}+1}$. 
There are three submodules in this layer. 

The first sub-module (denoted as $S$) is a linear regression part similar to that of FWFM, which models global bias of data and weight of categorical features. It operates on the categorical vectors, i.e. $\mathbf{h}_S=w_0 + \sum_{i=1}^{N}\sum_{j=1}^{K_i} w_{i}[j] x_{i}[j]$ where $w_{0}$ is bias term. Note that for ease of description, we let $\mathbf{W}_{S}$ denote the set of all $w_0\cup w_{i}$.

The second sub-module is called the MF module, which focuses on learning inter-field feature interactions between each pair of the hierarchical fields. It first operates element-wise product on all pairs of field-wise embedding vectors, i.e.  $\mathbf{h}_{MF}= \sum_{i=1}^{M} \sum_{j=i+1}^{M}  \mathbf{e}_i\odot \mathbf{e}_j r[i][j]$, where $r[i][j] \in \Real$ is a weight to model the interaction strength between field $i$ and $j$. We use $\odot$ to denote the element-wise product of two vectors, that is, $(\textbf{\textit{e}}_i \odot \textbf{\textit{e}}_j)[k] = \mathbf{e}_{i}[k]\mathbf{e}_{j}[k]$. %$(\textbf{\textit{v}}_i \odot \textbf{\textit{v}}_j)[k] = \mathbf{v}_{i}[k]\mathbf{v}_{j}[k]$. 

In real industrial system, the quantity of feature fields is usually 10 or more, for example there are 33 feature fields in our industrial dataset, but the hierarchical field number, $M$, is usually less than 4~\citep{liu2017pbodl,covington2016deep} for reducing computation and avoiding overfitting. This hierarchical field manner is inspired by YouTube, according to whether they describe properties of the item or properties of the user/context~\citep{covington2016deep}.
As shown in Figure~\ref{fig:overview}, there are 3 MF models to learn field-wise feature interactions for each pair of the hierarchical user, item and context fields, respectively.

The third sub-module is called the FM module, which focuses on learning intra-field feature interactions in each FM module. It first computes self element-wise product on each field embedding vectors, i.e. $\mathbf{hf}_{m} =\mathbf{e}_m\odot \mathbf{e}_m$. 
Similar operations are also conducted on each feature embedding vectors,  i.e. $\mathbf{ht}_{m} =\sum_{n, F(n)=m} \mathbf{e}_n \odot \mathbf{e}_n$. 
Finally, the two vectors are merged over all field-wise subtraction of the two intermediate vectors, i.e. $\mathbf{h}_{FM}= \sum_{m} (\mathbf{hf}_{m}-\mathbf{ht}_{m} ) r[m][m]$, where $r[m][m] \in \Real$ is a weight for each field $m$, discriminating the importance of each field that contribute to the final prediction. Note that for ease of description, we let $\mathbf{R}$ denote the set of all $r[i][j]\cup r[m][m]$.

Clearly, the output of MF module and FM module is a $K_e$-dimension vector that encodes the inter-field and intra-field feature interactions in the embedding space, respectively.

We concatenate the output of $S$ and the sum pooling of MF and FM module, i.e. $\mathbf{h}_{in}=[\mathbf{h}_S,\mathbf{h}_{MF} + \mathbf{h}_{FM}]$. 
Then $\mathbf{h}_{in}$ is fed to a hidden layer,  i.e. $\mathbf{h}_{FwBI}=\sigma( \mathbf{W}_{FwBI}^T \mathbf{h}_{in} )$. 
In practice, we use the ReLU as the active function $\sigma$. 

By leveraging multiple MF and FM modules to learn both the inter-field and intra-field feature interactions, we end up with more disentangled parameters and therefore with faster training. Furthermore, the parallel structures of FwBI allows the computational budget be distributed in a distributed environment. 

It is worth pointing out that the FwBI pooling layer is much more memory-efficient than FFM and furthermore, it can be efficiently computed in $O(MK_eN + K_eM^2)$. In real industrial system, the quantity of feature fields is usually 10 or more, for example there are 33 feature fields in our industrial dataset, but the hierarchical field number, $M$, is usually less than $4$~\citep{liu2017pbodl,covington2016deep}. There are $3$ hierarchical field, i.e. user, item and context, in our industrial dataset. When $M \ll N$, FwBI can be efficiently trained and serverd online in linear time, which is very attractive in industrial systems.
%output MF 32=32

\subsection{Relation to Previous CTR Prediction Systems}
A variety of shallow models can be expressed and generalized under the field-wise bi-interaction technique. 
To start with, we set the active function $\sigma$ as an identify function, the weight matrix $\mathbf{W}_{FwBI}^T=\mathbf{I}$, where $\mathbf{I}$ is a unit matrix. 
Since the embedding vectors $\mathbf{e}_m,\mathbf{e}_n$ are transformed from the original feature representations, $\mathbf{e}_m=\sum_{n|F(n)=m} \mathbf{e}_{n}$, $\mathbf{e}_n = \mathbf{V}_{n}\mathbf{x}_n$, we can re-write the computation in field-wise bi-interaction pooling layer as: 
\begin{equation}\label{equ:flen}%check for v/V
\begin{split}
&	\Phi_{FwBI}= \mathbf{W}_{FwBI}^T\Big[\underbrace{w_0+\sum_{i=1}^{N}\sum_{j=1}^{K_i} w_{i}[j] x_{i}[j] } _{\textit{S}} \\
&	 , \underbrace{\sum_{i=1}^{M} \sum_{j=i+1}^{M}\big[(\sum_{n,F(n)=i } \mathbf{V}_{i} \mathbf{x}_n)(\sum_{n, F(n)=j } \mathbf{V}_{j} \mathbf{x}_n) r[i][j]}_{\textit{MF}}\big] \\
&	 + \underbrace{ \sum_{m}^M [(\sum_{n, F(n)=m } \mathbf{V}_{n} \mathbf{x}_n)^2-( \sum_{n,F(n)=m} (\mathbf{V}_{n} \mathbf{x}_n)^2)] r[m][m]}_{\textit{FM}}\Big]
\end{split}
\end{equation}   
where we use the symbol $(\mathbf{Vx})^2$ to denote $\mathbf{Vx}\odot \mathbf{Vx}$. The first term in Equation~\ref{equ:flen} corresponds to the first sub-module which is based on single feature values. The second term corresponds to the sum pooling over the second sub-module, which resembles the matrix factorization form when written in feature representations, and the third sub-module, which resembles the factorization machine form.

Clearly, if there are only one field, i.e., $M=1$ and $r_{m, m}=\frac{1}{2}$, then we can exactly recover the FM model.
\begin{equation}\label{eq:fm}
\begin{split}
&	\Phi_{FwBI}=\Phi_{FM}\\
%&=\Big[w_0+\sum_{i=1}^Nw[i]x[i], \frac{1}{2}[(\sum_{i=1}^N v[i]x[i])^2-\sum_{i=1}^N((v[i]x[i])^2]\Big]
&=\Big[w_0 + \sum_{i=1}^{N}\sum_{j=1}^{K_i} w_{i}[j] x_{i}[j], \frac{1}{2}[(\sum_{i=1}^N \mathbf{V}_i\mathbf{x}_i)^2-\sum_{i=1}^N(\mathbf{V}_i\mathbf{x}_i)^2]\Big]
\end{split}
\end{equation}   

If there are $N$ fields, i.e., $M=N$, there is not any intra-field feature interactions, then we can recover the FwFM model.
\begin{equation}\label{eq:fwfm}
\begin{split}
&	\Phi_{FwBI}=\Phi_{FwFM}\\
%&=\Big[w_0+\sum_{i=1}^M w[i]x[i], \sum_{i=1}^M\sum_{j>i}^M(v[i]x[i])(v[j]x[j])r[i][j]\Big]
&=\Big[w_0 + \sum_{i=1}^{M}\sum_{j=1}^{K_i} w_{i}[j] x_{i}[j], \sum_{i=1}^M\sum_{j>i}^M(\mathbf{V}_i\mathbf{x}_i)(\mathbf{V}_j\mathbf{x}_j)r[i][j]\Big]
\end{split}
\end{equation}

%It is worthy to point out that FLEN adopts a smaller number of parameters in this layer. 
%The number of parameters in FMs is $n + nK$, where $n$ accounts for the weights for each feature in the linear part $\{w_i|i=1,...,n\}$ and $nK$ accounts for the embedding vectors for all the features $\{\textbf{\textit{v}}_i|i=1,...,m\}$. FwE use $m(m-1)/2$ additional parameters $\{r_{F_i, F_j} | i, j = 1,...,m\}$ for each field pair so that the total number of parameters of FwE is $n + nK + m(m-1)/2$. For FFMs, the number of parameters is $n + n(m-1)K$ since each feature has $m-1$ embedding vectors. Given that usually  $m\ll n$, the parameter number of FwE is comparable with that of FMs and significantly less than that of FFMs.

%\subsubsection{\textbf{Time Complexity Analysis}}
%\label{subsec:time_complexity}

\subsection{MLP component}\label{sec:MLP}
We employ an MLP component to capture non-linear, high-order feature interactions. 
The input is simply a concatenation of all field-wise embedding vectors, i.e. $\mathbf{h}_0=concat(\mathbf{e}_1,\cdots,\mathbf{e}_M)$. 
A stack of fully connected layers is constructed on the input $\mathbf{h}_0$. 
Formally, the definition of fully connected layers are as follows:
\begin{equation}
\begin{aligned}
	\mathbf{h}_1 &= \sigma_1 (\textbf{W}_1 \mathbf{h}_0 + \textbf{b}_1 ), \\
	\mathbf{h}_2 &= \sigma_2 (\textbf{W}_2 \mathbf{h}_1 + \textbf{b}_2), \\
	&\quad\quad ... ... \\
	\mathbf{h}_L &= \sigma_L (\textbf{W}_L \mathbf{h}_{L-1} + \textbf{b}_L),
\end{aligned}
\end{equation}
where $L$ denotes the number of hidden layers, $\textbf{W}_l$, $\textbf{b}_l$ and $\sigma_l$ denote the weight matrix, bias vector and activation function for the $l$-th layer, respectively. We use ReLU as the active function for each layer. Note that for ease of description, we let $\mathbf{W}_{MLP}$ denote the set of all $\textbf{W}_l$, and $\mathbf{b}_{MLP}$ denote the set of all $\textbf{b}_l$.

\subsection{Prediction Layer}\label{sec:prediction}

The output vector of the last hidden MLP layer $\textbf{h}_L$ is concatenated with the output vector of field-wise bi-interaction pooling layer $\Phi_{FwBI}$ to form $\mathbf{h}_F = concat(\mathbf{h}_{FwBI},\mathbf{h}_L)$. 
The concatenation $\mathbf{h}_F$ goes through one last hidden layer and is transformed to 
\begin{equation}
	z = \sigma(\mathbf{w}^T_F \mathbf{h}_F), 
\end{equation}
where vector $\mathbf{w}_F$ denotes the neuron weights of the final hidden layer. 

Finally, we apply a sigmoid layer to make predictions. 
\begin{equation}
	\sigma(\mathbf{z}) = \frac{1}{1+e^{-z}} 
\end{equation}

Our loss function is \textit{negative log-likelihood}, which is defined as follows:
\begin{equation}\label{eqa::loss}
\mathbb{L} = -\frac{1}{|\mathcal{X}|} \sum_{s=1}^{|\mathcal{X}|} (y^s log(\Phi(\mathbf{x}^s))+(1-y^s)log(1-\Phi(\mathbf{x}^s))),
\end{equation}
where $y^s \in \{0, 1\}$ as the label, $\Phi(\mathbf{x}^s)$ is the output of the network, representing the estimated probability of the instance $\mathbf{x}^s$ being clicked. The parameters to learn in our model are represented as $\mathbf{V}, \mathbf{W}=\{\mathbf{W}_{S},\mathbf{W}_{FwBI},\mathbf{W}_{MLP},\mathbf{w}_F\},\mathbf{b}_{MLP},\mathbf{R}$, which are updated via minimizing the total negative log-likelihood using gradient descent.

\section{Dicefactor: Dropout Method}
As noted in previous work~\citep{qu2018product}, FM may cause the coupled gradient issue because it uses the same latent vectors in different types of inter-field interactions, i.e.  two supposedly independent features are updated in the same direction during the gradient update process. It will damage the performance of the model. To solve this challenge, we propose a novel dropout method to decouple independent features, i.e., Dicefactor. 
Dicefactor is inspired by Dropout~\citep{srivastava2014dropout}. It randomly drops bi-linear paths (i.e. cross-feature edge in the FM module of field-wise bi-interaction pooling layer) to prevent a feature from adapting to other features. 

\begin{figure}[htbp]
\centering
  \begin{subfigure}{0.48\linewidth}
    \includegraphics[width=\linewidth]{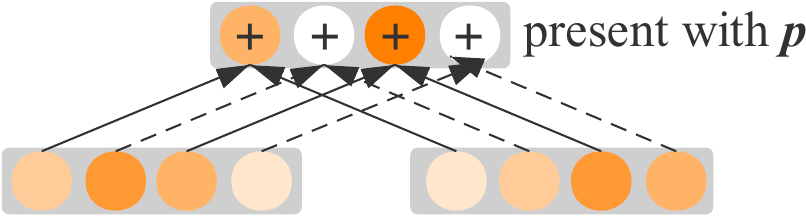}
    \caption{train phase}
    \label{fig::subfig::a}
  \end{subfigure}
  \begin{subfigure}{0.48\linewidth}
    \includegraphics[width=\linewidth]{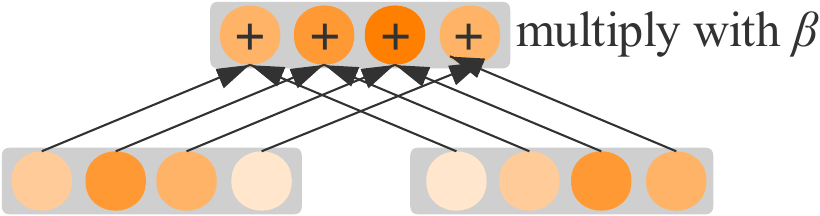}
    \caption{inference phase}
    \label{fig::subfig::b}
  \end{subfigure}
  \caption{Dicefactor keeps a bi-linear path with probability $\beta$ at train phase. At inference phase, every bi-linear path is kept and the output is multiplied by $\beta$.}
  \label{fig::subfig}
\end{figure}

We first reformulate Eq.~(\ref{eq:fm}) as:
\begin{equation}\label{eq:expand}
\begin{aligned}
\Phi_{FM}=\Big[w_0 + \sum_{i=1}^{N}\sum_{j=1}^{K_i} w_{i}[j] x_{i}[j], \sum_{i=1}^M \sum_{j=1\& i \neq j}^M \mathbf{e}_i\odot \mathbf{e}_j\Big], 
%\Big[w_0+\sum_{i=1}^Nw[i]x[i] , \sum_{i=1}^M \sum_{j=1\& i \neq j}^M \mathbf{e}_i\odot \mathbf{e}_j\Big], 
\end{aligned}
\end{equation}
where $\mathbf{e}_i$ represents the $i$-th field's embedding vector. Based on Eq.~(\ref{eq:expand}), Fig.~\ref{fig::subfig} shows the expanding structure of the bi-linear interaction in two field embedding vectors. There are $K_{e}$ bi-linear paths, each bi-linear path connects corresponding elements in two embedding vectors. The key idea of DiceFactor is to randomly drop the bi-linear paths during the training. This partly prevents $\mathbf{e}_i$ from co-adapting to $\mathbf{e}_j$, i.e., $\mathbf{e}_i$ is updated through the direction of $\mathbf{e}_j$. 

In our implementation, each factor is retained with a predefined probability $\beta$ during training. With the DiceFactor, the formulation of bi-linear interaction of FM part in the training becomes:
\begin{equation}\label{eq:drop}
\begin{aligned}
\Phi_{FM}=\Big[w_0 + \sum_{i=1}^{N}\sum_{j=1}^{K_i} w_{i}[j] x_{i}[j], \sum_{i=1}^M \sum_{j=1 \&i \neq j}^M \mathbf{p}  \mathbf{e}_i\odot \mathbf{e}_j\Big],
%\Big[w_0+\sum_{i=1}^N w[i]x[i] , \sum_{i=1}^M \sum_{j=1 \&i \neq j}^M \mathbf{p}  \mathbf{e}_i\odot \mathbf{e}_j\Big],
\end{aligned}
\end{equation}
where $\mathbf{p} \in \Real^{K_{e}}$ and $p[i] \sim Bernoulli(\beta)$. With the DiceFactor, the network can be seen as a set of $2^{K_{e}}$ thinned networks with shared weights. In each iteration, one thinned network is sampled randomly and trained by back-propagation as shown in Fig.~\ref{fig::subfig::a}.

For inference, instead of explicitly averaging the outputs from all $2^{K_{e}}$ thinned networks, we use the approximate ``Mean Network'' scheme in \citep{srivastava2014dropout}. As shown in Fig.~\ref{fig::subfig::b}, 
%each factor term $x_i\textbf{\textit{v}}_i\odot x_j\textbf{\textit{v}}_j$ is multiplied by $\beta$ at inference phase:
each factor term $\mathbf{e}_i\odot \mathbf{e}_j$ is multiplied by $\beta$ at inference phase:
\begin{equation}\label{eq:drop_test}
\begin{aligned}
\Phi_{FM}=\Big[w_0 + \sum_{i=1}^{N}\sum_{j=1}^{K_i} w_{i}[j] x_{i}[j], \sum_{i=1}^M \sum_{j=1 \&i \neq j}^M \beta \mathbf{e}_i\odot \mathbf{e}_j\Big]
%\Big[w_0+\sum_{i=1}^Nw[i]x[i] , \sum_{i}^M \sum_{j \&i \neq j}^M \beta \mathbf{e}_i\odot \mathbf{e}_j\Big]
\end{aligned}
\end{equation}

In this way, the output of each neuron at inference phase is the same as the expectation of output of $2^{K_{e}}$ different networks at train phase.

\section{Offline Model Evaluation}

To assess the validity of our CTR prediction models, we run a traditional offline evaluation based on a well-known public benchmark and an industrial dataset.

\textbf{Avazu}\footnote{https://www.kaggle.com/c/avazu-ctr-prediction} The first dataset we adopt is originally used in the Kaggle CTR prediction competition. It contains users' mobile behaviors, i.e. whether a displayed mobile ad is clicked by a user. It has $23$ feature fields spanning from user/device features to ad attributes. The data set is collected during a time span of $10$ days. We use $9$ days of clicks for training and the last $1$ day of data for test.

Note that in this paper we do not use another well-known benchmark, i.e., Criteo dataset, because the semantic of its features is undisclosed. We do not known which hierarchical field each feature belongs to.

%industrial
\textbf{Meitu} The second dataset  used to run the experiments is a uniformly generated sample of our internal historical data. As a training set we extract a sample of $2$ million items shown during a seven-day time period from 2019-07-26 to 2019-08-01 to users of a photo and video-sharing social networking app, namely Meitu.  We collect over $1.5$ billion users' records. The test set contains a sample of $1$ million items shown the next day, i.e. 2019-08-02. There are around $5$ million features (e.g., user age, clicked feed ids,  and etc) organized in hierarchical field manner, including ``user", ``item" and ``context".  Features used in our system are described in Table \ref{tab:feature}.

\begin{table}[htp]
\caption{Example features in Meitu dataset.}
\begin{center}
 \resizebox{\columnwidth}{!}{%
\begin{tabular}{llccc}
\toprule
Field & Feature & Dimemsionality & Type & AverageNonzero Ids per Instance\\ \toprule
\multirow{5}{7em}{User Field }
& gender & 2 &  one-hot & 1  \\ \cline{2-5}
& age & $\sim 10$ &  one-hot & 1  \\ \cline{2-5}
& clicked\_feed\_ids& $\sim 10^7$ &  multi-hot & $\sim 10^2$  \\ \cline{2-5}
& liked\_feed\_ids& $\sim 10^5$ &  multi-hot & $\sim 10^2$  \\ %\cline{2-5}
%& ... & ... & ... & ...  \\ 
\midrule
\multirow{5}{7em}{Item Field}
& feed\_id & $\sim 10^7$ &  one-hot & 1  \\ \cline{2-5}
& tags& $\sim 10^3$ &  multi-hot & $\sim 10^1$  \\ \cline{2-5}
& clicked\_rate& 10 &  one-hot & 10 \\ \cline{2-5}
& liked\_rate& 10 &  one-hot & 10 \\ \cline{2-5}
& author\_id & $\sim 10^7$ &  one-hot & 1  \\ \cline{2-5}
& author\_gender & 2 &  one-hot & 1  \\ \cline{2-5}
& author\_age & $\sim 10$ &  one-hot & 1  \\ \cline{2-5}
& author\_clicked\_rate& 10 &  one-hot & 10 \\ \cline{2-5}
& author\_liked\_rate& 10 &  one-hot & 10 \\ %\cline{2-5}
%& ... & ... & ... & ...  \\ 
\midrule
\multirow{4}{7em}{Context Field }
& network\_type & $\sim 4$ &  one-hot & 1  \\ \cline{2-5}
& time & $\sim 10$ &  one-hot & 1  \\ \cline{2-5}
& brand & $\sim 15$ &  one-hot & 1  \\ %\cline{2-5}
%& ... & ... & ... & ...  \\ 
\bottomrule
\end{tabular}}
\end{center}
\label{tab:feature}
\end{table}%

The statistics of the data sets are summarized in Table~\ref{tab::dataset}.

\begin{table}[t]
\centering\caption{Statistics of evaluation data sets.}
\begin{tabular}{| l | c | c | c | c} \hline
Data & \#Samples & \#Fields & \#Features (Sparse)  \\ \hline
Avazu & 40,428,967 & 23 & 1,544,488 \\ \hline
Meitu & 1,508,149,301 & 33 & 5,476,029\\ \hline
\end{tabular}\label{tab::dataset}
\end{table}

\subsection{Competitors}

We compare FLEN with $7$ state-of-the-art models. \\
(1) \textbf{FFM}~\citep{juan2017field}: a shallow model in which the prediction is aggregated over inner products of feature vectors. It represents a feature by several separate vectors, depending on the multiplying feature field. \\
(2) \textbf{FwFM}~\citep{pan2018field}: a shallow model which also explicitly aggregates over feature products. Interaction weights are assigned for each field pair. \\
(3) \textbf{DCN}~\citep{wang2017deep}: a deep model that takes the outer product of feature vectors at bit-wise level to a feed-forward neural network. \\
(4) \textbf{DeepFM}~\citep{guo2017deepfm}: a deep model that consists of a wide component that models factorization machine and a deep component. \\
(5) \textbf{NFM}~\citep{he2017neural}: a deep model which stacks MLP on top of a bi-interaction pooling layer. \\
(6) \textbf{xDeepFM}~\citep{lian2018xdeepfm}: a deep model that explicitly generates features with a compressed interaction network. \\
(7) \textbf{NFFM}\citep{Yang2019Operation}: a deep model which learns different feature representations for convolutional operations and product operations.

All methods are implemented in TensorFlow\footnote{Codes are available at https://github.com/aimetrics/jarvis}.
We use an embedding dimension of 32 and batch size of 512 for all compared methods. Hidden units $d'$ are set to 64, 32. We use AdaGrad~\citep{duchi2011adaptive} to optimize all deep neural network-based models. DCN has two interaction layers, following by two feed-forward layers. We use one hidden layer of size 200 on top of Bi-Interaction layer for NFM as recommended by their paper. 
We use $2$ layers with size $(64,32)$ in the MLP component of FLEN. 
All experiments are executed on one NVIDIA TITAN Xp Card with 128G memory.

\subsection{Evaluation Metrics}

We use two commonly adopted evaluation metrics.

\textbf{AUC} Area Under the ROC Curve (AUC) measures the probability that a CTR predictor will assign a higher score to a randomly chosen positive item than a randomly chosen negative item. A higher AUC indicates a better performance.

\textbf{Logloss} Since all models attempt to minimize the \textit{Logloss} defined by Equation~\ref{eqa::loss}, we use it as a straightforward metric.

It is now generally accepted that increase in terms of AUC and Logloss at $0.001$-level is significant~\citep{cheng2016wide,guo2017deepfm,wang2017deep}.

\subsection{Comparative Performance}\label{sec::result}
We report the AUC and Logloss performance of different models in Table~\ref{tab:results}. 
We distinguish the original FLEN (denoted as FLEN) and the model with Dicefactor implementation (denoted as FLEN+D). 
We can see that on both datasets, FLEN has achieved the best performance in terms of AUC and Logloss. 
We point out that FLEN has impressively boosted the AUC performance of the best competitor (i.e. NFFM) by $0.002$ on the industrial dataset, which validates the superiority of FLEN on large-scale CTR systems. 
We also observe that Dicefactor further significantly enhances AUC performance of FLEN on both datasets.   

Furthermore, we can find an interesting observation: leveraging field information makes modeling feature interactions more precisely. This observation is derived from the fact that by exploiting field information, FLEN, NFFM and xDeepFM perform better than NFM, DeepFM and DCN do on the Avazu and Meitu datasets. 
This phenomenon can be found in more literature, including FFM~\citep{juan2017field} and FwFM~\citep{pan2018field}.

\begin{table}[!htbp]
\centering\caption{AUC and Logloss performance of different models}
\begin{tabular}{|c|c|c|c|c|} \hline
\multirow{2}{*}{Model} & \multicolumn{2}{|c|}{Avazu} & \multicolumn{2}{|c|}{Meitu} \\ \cline{2-5}
& AUC & Logloss & AUC & Logloss \\ \hline
FFM  & 0.7400 & 0.3994 & 0.6300 & 0.5625 \\
FwFM  & 0.7406 & 0.3988 & 0.6306 & 0.5621 \\
DCN  & 0.7421 & 0.3981 & 0.6337 & 0.5606 \\
DeepFM  & 0.7438 & 0.3982 & 0.6329 & 0.5612 \\
NFM  & 0.7449 & 0.3973 & 0.6359 & 0.5596 \\
xDeepFM & 0.7509 & 0.3947 & 0.6440 & 0.5576 \\
NFFM & 0.7513 & 0.3945 & 0.6443 & 0.5565 \\
FLEN  & \textbf{0.7519} & \textbf{0.3944} & \textbf{0.6463} & \textbf{0.5558}\\ 
FLEN+D & \textbf{0.7528} & \textbf{0.3944} & \textbf{0.6475} & \textbf{0.5554}\\\hline
\end{tabular}
\label{tab:results}
\end{table}

\subsection{Memory Consumption and Running Time}

To illustrate the scalability of FLEN, we first compare the model complexity and actual parameter size on Avazu dataset of each method in Table~\ref{tab:complexity} . 
For a fair comparison, in computing parameter size, we assume the number of deep layers denoted as $H=3$, the number of hidden layers denoted as $L=3$ and embedding size $K_e=32$. We do not take into account parameters for the feed-forward neural network. 
The number of features is $N$=1,544,448 and the number of fields is $M=23$ on Avazu dataset.
%Note that if we explicitly model all possible pairwise feature interactions in Internal dataset, the dimensionality scales up to $10^9$.
We can see that FLEN is one of the models that make use of the smallest number of parameters. 
%(Ryan: not this reason.).This is because FLEN organizes features in a hierarchical field manner, thus the parameter size depends on the number of hierarchical fields $M$, which is usually less than $4$~\citep{liu2017pbodl,covington2016deep} in practice. This hierarchical field manner is inspired by YouTube, according to whether they describe properties of the item or properties of the user/context~\citep{covington2016deep}.

\begin{table}[!htbp]
\centering\caption{Model complexity and parameter size in Avazu dataset for different models}. 
\resizebox{\columnwidth}{!}{
\begin{tabular}{|c|c|c|} 
\hline
Model & Model Complexity & Parameter Size  \\ \hline
FFM & $O(NMK_e)$ & $1.14 \times 10^9$ \\
FwFM & $O(NK_e+ M^2)$ & $4.94 \times 10^7$\\
DCN & $O(NK_e + MK_eL + MK_eH)$ & $4.94 \times 10^7$\\
DeepFM & $O(NK_e + MK_eH)$ & $4.94 \times 10^7$\\
NFM & $O(NK_e + K_eH)$ & $4.94 \times 10^7$\\
xDeepFM & $O(NK_e + MH^2L + MK_eH)$ & $4.94 \times 10^7$\\
NFFM & $O(NMK_e)$ & $1.14 \times 10^9$ \\
FLEN & $O(NK_e + M^2 + MK_eH)$ & $4.94 \times 10^7$\\ \hline
\end{tabular}\label{tab:complexity}}
\end{table}

\begin{figure}[ht]
	\centering
	\begin{subfigure}[b]{\columnwidth}
		\centering
		\includegraphics[width=\columnwidth]{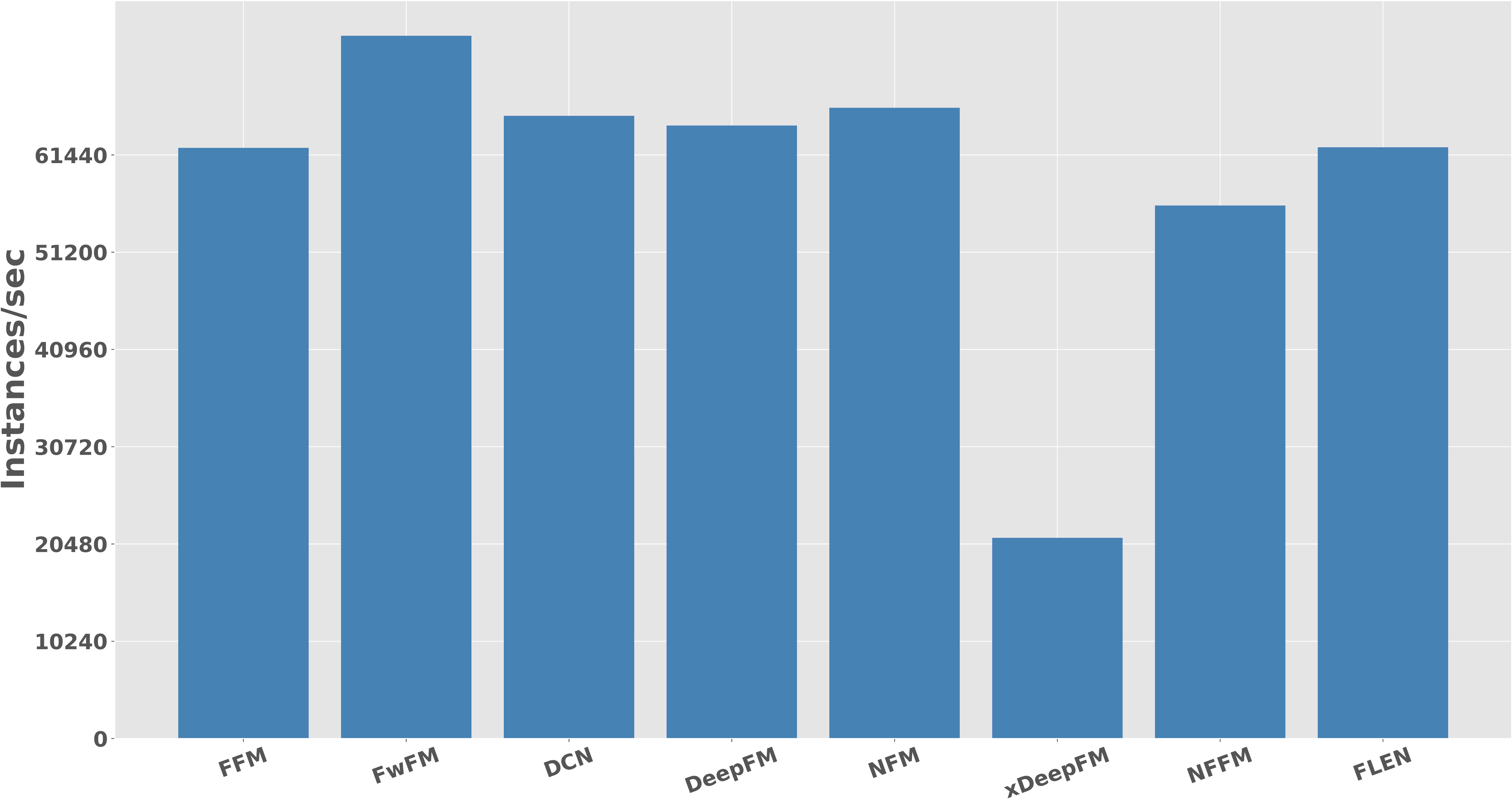}
		\caption{Avazu}
		\label{fig:runtimeavazu}
	\end{subfigure} \hspace{-7pt}
	\begin{subfigure}[b]{\columnwidth}
		\centering
		\includegraphics[width=\columnwidth]{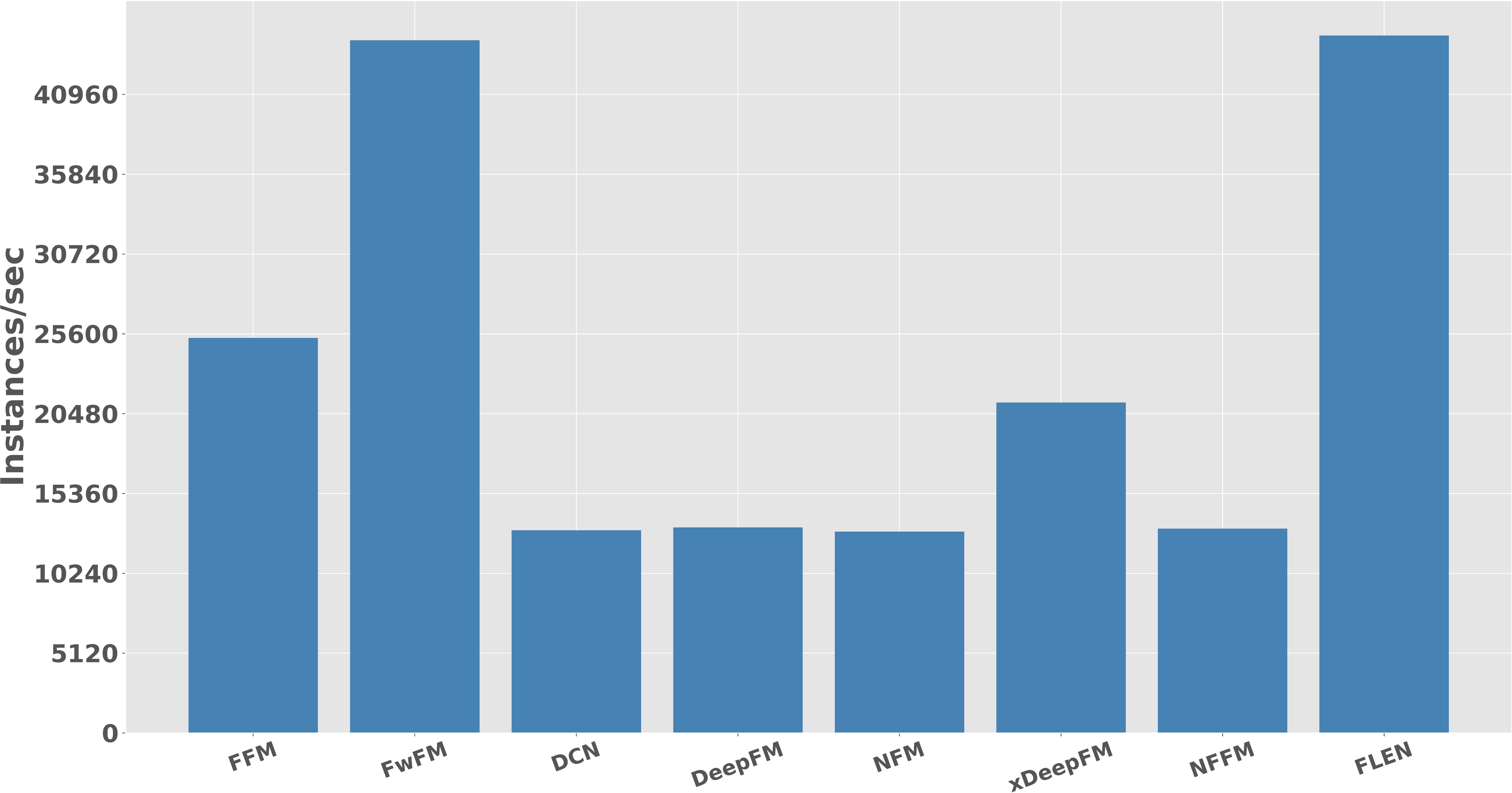}
		\caption{Meitu}
		\label{fig:runtimemeitu}
	\end{subfigure}  \hspace{-7pt}
	\caption{Number of instances per second processed by different models }
	\label{fig:runtime}
\end{figure}

For a detailed study, we report the number of instances being processed by different models per second on the two datasets.  
As shown in Figure~\ref{fig:runtime}, FLEN operates on the most instances on Meitu dataset. On Avazu dataset, FLEN is comparable with other state-of-the-art methods. 
But the high efficiency makes FLEN applicable in real industrial systems to handle large scale and high-dimensional data.

\begin{figure}[htbp]
	\centering
	\begin{subfigure}[b]{\columnwidth}
		\centering
		\includegraphics[width=\textwidth]{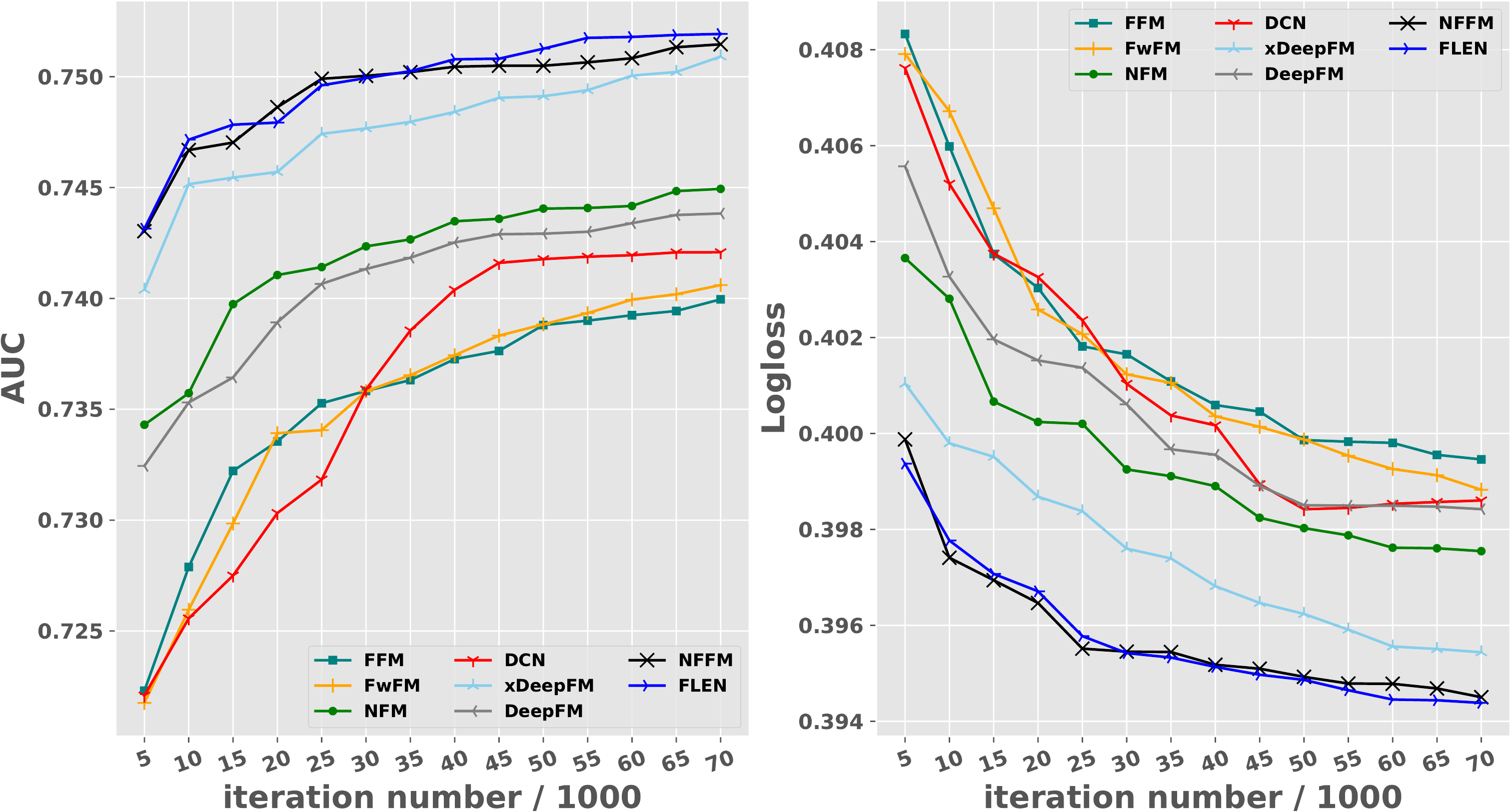}
		\caption{Avazu}
		\label{fig:convergeavazu}
	\end{subfigure} \hspace{-7pt}
	
	\begin{subfigure}[b]{\columnwidth}
		\centering
		\includegraphics[width=\textwidth]{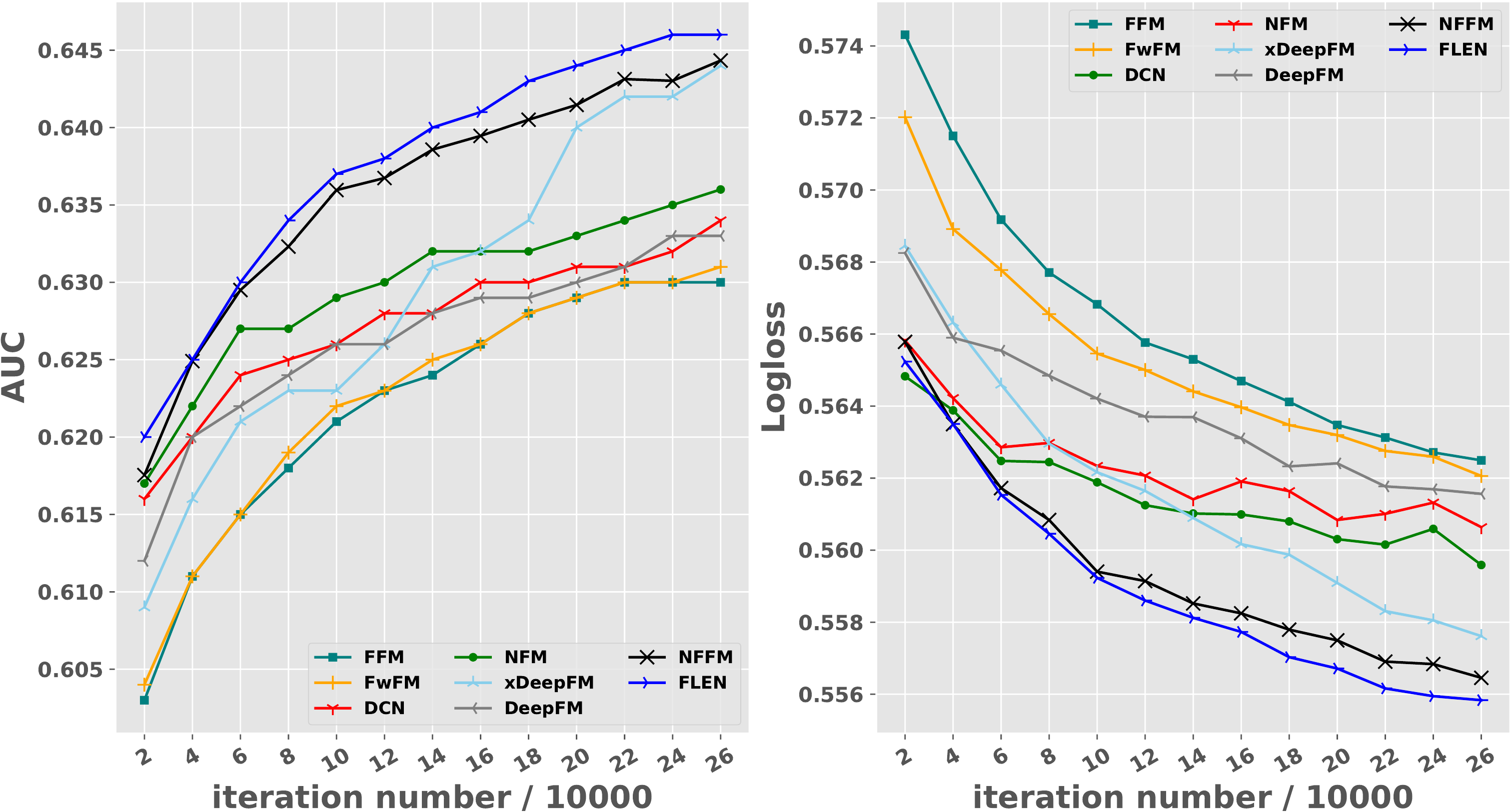}
		\caption{Meitu}
		\label{fig:convergemeitu}
	\end{subfigure}  \hspace{-7pt}
	\caption{AUC and Logloss change in training time.}
	\label{fig:converge}
\end{figure}

We keep track of AUC and Logloss during the training process after each training ``epoch" (i.e., 5,000 iterations through all the training data on Avazu and 20,000 iterations on Meitu ).
As shown in Figure~\ref{fig:converge}, FLEN obtains the best AUC (i.e. highest) and Logloss (i.e. lowest) on both datasets in each iteration. 
Furthermore, we point out that although NFFM (which is the best competitor) is close with FLEN, FLEN achieves faster convergence towards optimization. For example, FLEN has a sharper increase of AUC and more steep decrease of Logloss on Meitu dataset. 
Thus FLEN requires less training time than NFFM, which is desirable in real-world production systems. 

\subsection{Impact of Parameters}
\begin{figure}
\centering\includegraphics[width=\columnwidth]{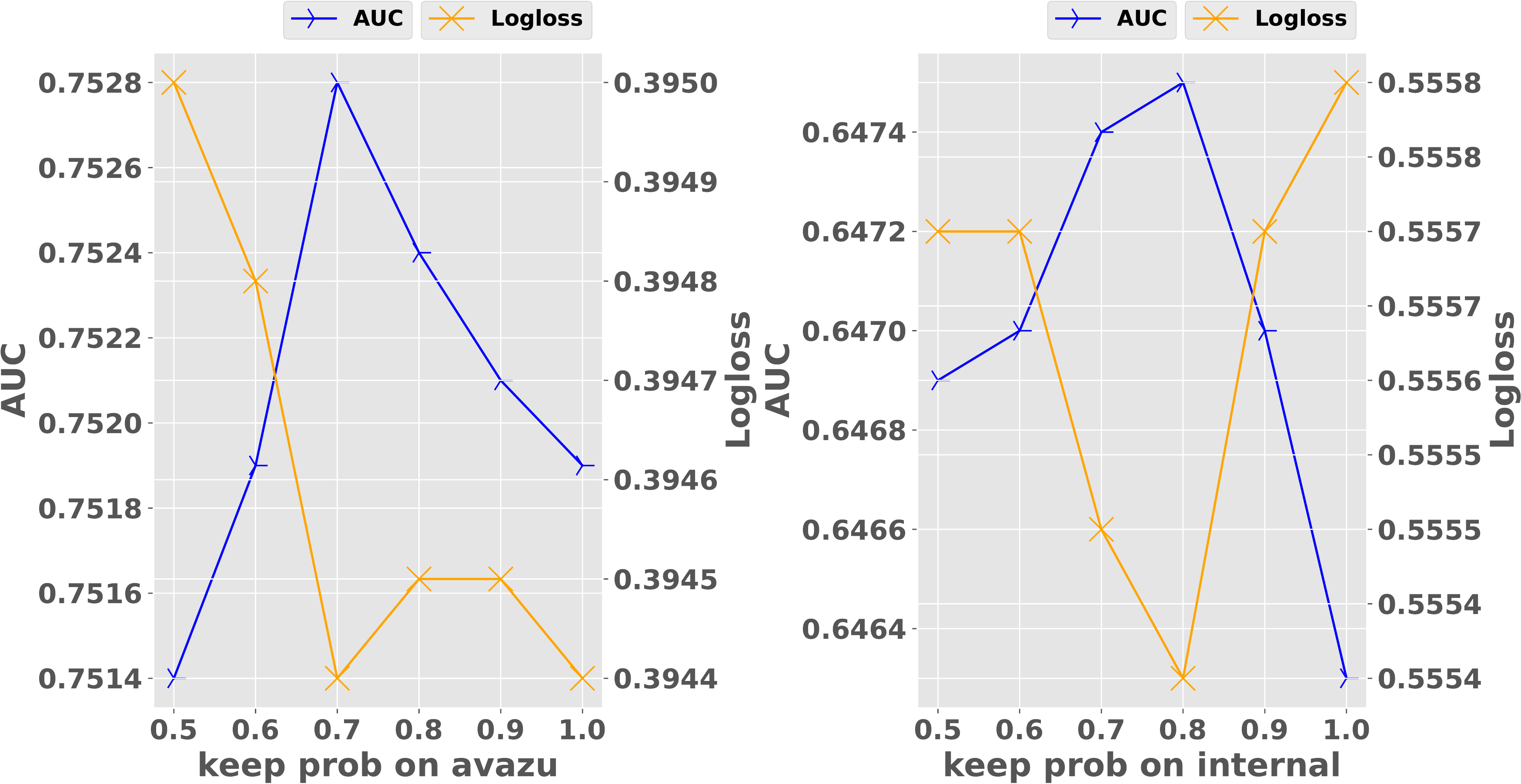}
\caption{AUC and logloss with different keep probability of dicefactor on two datasets. }
\label{fig:dicefactor}
\end{figure}

We analyze the impact of an important parameter of Dicefactor, the probability $\beta$ to keep bi-linear paths. 
We empirically find that a small value $\beta \in (0,0.5)$ leads to poor performance.
Henceforth, we set $\beta=0.5,0.6,0.7,0.8,0.9,1.0$ respectively and report the AUC and Logloss results.  
As shown in Figure~\ref{fig:dicefactor}, performance of FLEN is affected by $\beta$, the keep probability. 
The best parameter settings for AUC and Logloss are consistent. For example, best AUC and Logloss are both obtained at $\beta=0.7$ on Avazu dataset. On Meitu dataset, the best $\beta=0.8$. 

\section{Online Evaluation}

To measure the impact that FLEN has on users we conduct an online evaluation through a 7-day A/B testing on Meitu app. %a mobile photo-centered app.
We split $10\%$ of the incoming traffic as experiment group, the rest as control group. 
The items of the control group in the A/B testing period are provided by the previous version of online ranking system, which is based on NFM. 
The items offered to the experiment group are items that are predicted by FLEN , i.e. we deliver top $12$ items with highest prediction score by FLEN. 

\begin{figure}[htbp]
\centering\includegraphics[width=\columnwidth]{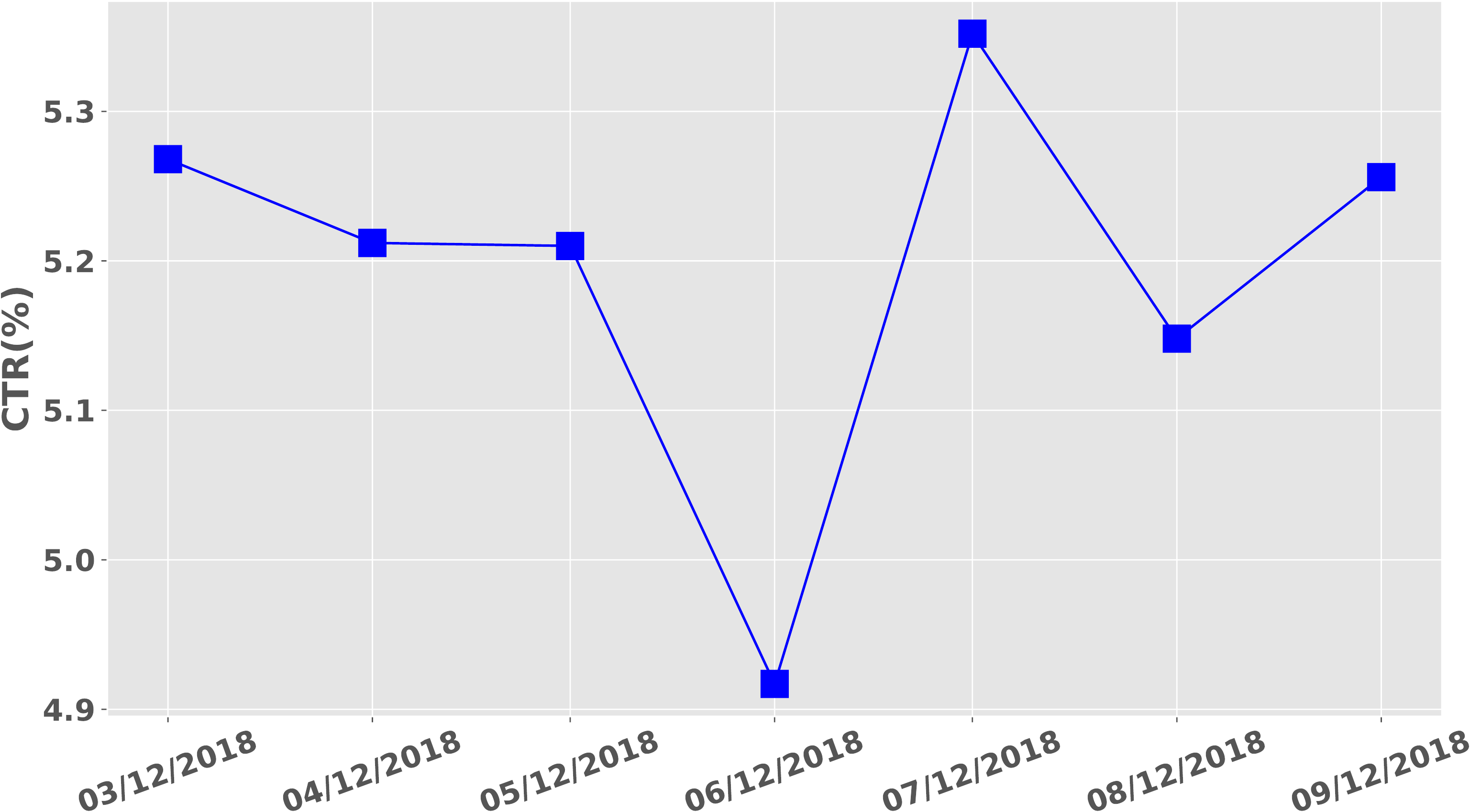}
\caption{CTR increase during our online A/B testing period.}
\label{fig:online}
\end{figure}

We report the CTR different during the A/B testing period in Figure~\ref{fig:online}. 
We observe stable and significant increase of CTR during the A/B testing period. 
The minimal CTR increase is above $4.9\%$. The mean CTR improvement over seven days is $5.195\%$ with variance $0.282\%$.

\section{Conclusion}
In this paper, we describe the FLEN model that is deployed in the online recommender systems in Meitu, serving the main traffic. %our large-scale mobile photo-centric app.  
FLEN has obtained significant performance increase by exploiting field information with an acceptable memory usage and computing latency in the real-time serving system. 
As future work, we plan to explore the usage of the attention mechanism to attend to important field embedding. Furthermore, we are interested in extending FLEN to  multi-task learning, i.e. predict conversions-rate and click-through-rate simultaneously. 

\begin{acks}
The authors thank the Runquan Xie for the valuable comments, which are beneficial to the authors’ thoughts on recommender systems and the revision of the paper.
The authors also thank Haoxuan Huang for his discussions and help in the extensive experiments at Meitu. Chen Lin is supported by the National Natural Science Foundation of China Nos. 61972328. Dugang Liu is supported by the National Natural Science Foundation of China Nos. 61872249, 61836005 and 61672358. 
\end{acks}

\end{document}